\documentclass{WileyMSP-template}

\usepackage{amsmath}
\usepackage{ulem}
\usepackage{graphicx, subfigure}
\usepackage{pslatex}
\usepackage{relsize}

\usepackage{bm}
\usepackage{verbatim}
\usepackage{float}
\usepackage{appendix}
\usepackage{booktabs}
\usepackage{xcolor}
\usepackage{soul}
\usepackage{ulem}
\usepackage[utf8]{inputenc} 
\usepackage[T1]{fontenc}
\DeclareUnicodeCharacter{2212}{-}
\DeclareUnicodeCharacter{0105}{\c a}

\usepackage[dvipsnames]{xcolor}
\usepackage[numbers,super,square,sort&compress ]{natbib}
\begin{document}
\pagestyle{fancy}
\rhead{\includegraphics[width=2.5cm]{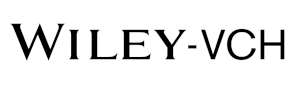}}

\title{
Development of a magnetic interatomic potential for cubic antiferromagnets: the case of NiO}

\maketitle

\author{Ievgeniia Korniienko}
\author{Pablo Nieves}
\author{Jakub Sebesta}
\author{Roberto Iglesias}
\author{Dominik Legut*}

\dedication{
}

\begin{affiliations}

Dr. I. Korniienko, Dr. J. Sebesta\\
IT4Innovations\\ 
V\v{S}B - Technical University of Ostrava\\
17. listopadu 2172/15, 70800 Ostrava-Poruba, Czech Republic\\

Dr. P. Nieves\\
Departamento de F\'{i}sica, Universidad de Oviedo\\
C. Leopoldo Calvo Sotelo, 18, 33007, Oviedo, Spain\\

Prof. R. Iglesias\\
Departamento de F\'{i}sica, Universidad de Oviedo\\
C. Leopoldo Calvo Sotelo, 18, 33007, Oviedo, Spain\\
ASturias RAw Materials Institute (ASRAM), Universidad de Oviedo\\
C. Gonzalo Gutiérrez Quirós, s/n, 33600, Mieres, Spain\\

Dr. D. Legut\\
Department of Condensed Matter Physics\\
Faculty of Mathematics and Physics\\
Charles University\\
Ke Karlovu 3, 121 16 Prague 2, Czech Republic\\
IT4Innovations\\ 
V\v{S}B - Technical University of Ostrava\\ 17. listopadu 2172/15, 70800 Ostrava-Poruba, Czech Republic\\
dominik.legut@matfyz.cuni.cz

\end{affiliations}


\begin{abstract}
 
Interatomic potentials are essential for molecular dynamics simulations of magnetic materials, yet incorporating magnetic features into potentials for complex antiferromagnets remains challenging. Nickel oxide (NiO), a prototypical cubic antiferromagnet, exemplifies this difficulty. Here we develop a methodology to integrate magnetic properties into interatomic potentials for cubic antiferromagnets 
by adding a magnetic Hamiltonian which includes both the Heisenberg exchange and N\'{e}el model.
We apply this approach to NiO by constructing two potentials: one based on the Born model of ionic solids and another using a reference-free modified embedded atom method. Both potentials include magnetoelastic interactions and are validated against Density Functional Theory calculations, showing excellent agreement in mechanical and magnetic properties at zero temperature. These models enable large-scale simulations of magnetoelastic phenomena in antiferromagnets and open avenues for molecular dynamics studies involving coupled electric and magnetic fields in metal oxides.
\end{abstract}

\section{Introduction}
Magnetoelastic coupling leads to a number of phenomena that are interesting from both a purely scientific and an applied point of view -- \textit{e.g.}  Joule magnetostriction, Villari effect, $\Delta E$ effect, magnetically induced changes in the elasticity, magnetovolume effect, Wiedemann effect, Matteuci effect, Nagaoka-Honda effect.\cite{booktremolet} They can be applied in various ways from, for example, use in magnetostriction-based sensors and actuators\cite{doi:https://doi.org/10.1002/0471216275.esm051, doi:10.1177/1045389X06072358, Hathaway_Clark_1993} to, on the contrary, the use of materials with Invar-like behavior,\cite{SAHOO20212242} where thermal expansion is compensated by bulk magnetostriction over a broad temperature range.
Moreover, understanding the  magnon−phonon coupling  in magnetic materials is important for developing viable quantum technologies.\cite{doi:10.1021/acs.nanolett.3c05070}

In compensated antiferromagnets (AFM) magnetoelastic interaction is known to be a possible source of equilibrium domain structure,\cite{Helen_Gomonay_2002, 10.1063/1.2008141} since it stands for a primary factor governing the width, internal structure, and interaction between domain walls\cite{PhysRevLett.91.237205}. Moreover, magnetoelasticity is responsible for shape effects,\cite{ Kornienko-2005} acoustic excitation of antiferromagnetic spin waves,\cite{Zhang_2022} optically driven magnon-phonon Fermi  resonance,\cite{Metzger2024} and represents a factor  affecting the propagation of surface acoustic waves,\cite{doi:10.1021/acs.nanolett.3c05070} etc.
All in all, many effects in AFM cannot be adequately studied if a proper description of the magnetoelastic interaction is missing. Thereby, the development of accurate numerical spin-lattice models, including  magnetic and elastic degrees of freedom, as well as their mutual interplay, becomes necessary.
Such models based on the combination of classical spin and molecular dynamics (SD-MD) have already been formulated and successfully validated for the case of cubic ferromagnets (FM).\cite{PhysRevB.78.024434,MA2016350, PhysRevB.103.094437, Nieves-PhysRevB.105.134430, PhysRevB.99.104302, PhysRevB.103.054439, Pankratova2024, PhysRevMaterials.9.024409,Korniienko-2024}
In particular, spin-lattice simulation is able to show
realistic magnon-phonon behavior of the FM system in the region close to resonance where analytical formulas of linear theory of magnetoelasticity fail.\cite{KORNIIENKO2025108264} 
However, for the case of more complex systems such as AFM, the development of accurate numerical models is still an ongoing process.
This situation is caused by the fact that the simulation of magnetoelastic effects requires the model to correctly reproduce both magnetic dynamics under the condition of variable distances between the magnetic moments of atoms and elastic properties (which are given by the interatomic potential) in the presence of magnetic interactions. In the case of the AFM, both tasks become challenging. Thus, only a few interatomic potentials for the room-temperature oxide AFM NiO, which is often considered as a prototypical AFM material with a simple magnetic structure and important applications, are available.\cite{Lewis_1985, Fisher-potential, 10.1063/5.0246100} 
These potentials include charge, which allows to study the material response to electric fields in molecular dynamics simulations, but none of them include magnetic interactions. There are advantages in having interatomic potentials sensitive to both electric and magnetic fields, as for example a correct description of the response  to both components of electromagnetic 
radiation, that could be exploited in THz range based applications, modeling magnetoelectric effects,\cite{SHI2021113652} etc. 


In this work, we propose a  methodology for developing interatomic potentials in cubic AFM materials capable of describing their magnetic properties using molecular dynamics simulations, and we apply it to the case of NiO. 

\section{Methodology}
\label{section:Methodology}

\subsection{Spin-lattice Hamiltonian}
\label{section:Hamiltonian}

For the atomistic spin-lattice simulations of an AFM,  we consider the following Hamiltonian 
\begin{equation}
\begin{aligned}
\mathcal{H}_{sl}(\textbf{r},\textbf{p},\textbf{s}) & = \sum_{i=1}^N\frac{\vert\textbf{p}_i\vert^2}{2m_i}+\sum_{i,j=1}^N\mathcal{V}(r_{ij})+\mathcal{V}_{mag}(\textbf{r},\textbf{s}),
\label{eq:Ham_tot}
\end{aligned}
\end{equation}
where $\textbf{r}_i$, $\textbf{p}_i$, $\textbf{s}_i$, and $m_i$ stand for the position, momentum, normalized magnetic moment and mass of each atom $i$ in the system, respectively, $\mathcal{V}(r_{ij})=\mathcal{V}(\vert \textbf{r}_i-\textbf{r}_j\vert)$ is the non-magnetic part of the interatomic potential energy and $N$ is the total number of atoms in the system with total volume $V$.

The magnetic part of the interatomic potential  $\mathcal{V}_{mag}$ corresponds to a magnetic Hamiltonian $\mathcal{H}_{mag}$ that  includes the exchange interaction, the N\'{e}el interaction $\mathcal{H}_{N\acute{e}el}$ and the Zeeman term: 
\begin{equation}
\begin{aligned}
\mathcal{V}_{mag}(\textbf{r},\textbf{s}) & =\mathcal{H}_{mag}(\textbf{r},\textbf{s})  = -\frac{1}{2}\sum_{i,j=1,i\neq j}^N J(r_{ij})\textbf{s}_i \textbf{s}_j+ \mathcal{H}_{N\acute{e}el}(\textbf{r},\textbf{s})\\
&-\mu_0 \sum_{i=1}^N \mu_i \textbf{H}\textbf{s}_i,
\label{eq:Ham_mag}
\end{aligned}
\end{equation}
where  $\mu_i$ is the atomic magnetic moment, $\mu_0$ is the vacuum permeability, $\textbf{H}$ is the external magnetic field, and $J(r_{ij})$ is the exchange parameter.


The magnetic anisotropic effects can be included in the spin-lattice model by adding the N\'{e}el interaction to the magnetic 
interaction potential, (\textbf{Equation \ref{eq:Ham_mag}})
\cite{PhysRevB.103.094437} through a two-ion Hamiltonian\cite{booktremolet}
\begin{equation}
\begin{aligned}
\mathcal{H}_{N\acute{e}el} & =   -\frac{1}{2}\sum_{i,j=1}^{N} \lbrace g(r_{ij})+ l_1(r_{ij})\left[ (\textbf{e}_{ij}\textbf{s}_i)(\textbf{e}_{ij}\textbf{s}_j)-\frac{\textbf{s}_i\textbf{s}_j}{3}\right]  \\
& + q_1(r_{ij})\left[ (\textbf{e}_{ij}\textbf{s}_i)^2-\frac{\textbf{s}_i\textbf{s}_j}{3}\right]\left[ (\textbf{e}_{ij}\textbf{s}_j)^2-\frac{\textbf{s}_i\textbf{s}_j}{3}\right]   \\
& + q_2(r_{ij})\left[ (\textbf{e}_{ij}\textbf{s}_i)(\textbf{e}_{ij}\textbf{s}_j)^3+(\textbf{e}_{ij}\textbf{s}_j)(\textbf{e}_{ij}\textbf{s}_i)^3\right]  \rbrace,
\label{eq:Neel_energy}
\end{aligned}
\end{equation}
where $\textbf{e}_{ij}=\textbf{r}_{ij}/r_{ij}$, and 
\begin{equation}
\begin{aligned}
l_1(r_{ij}) & = l(r_{ij})+\frac{12}{35}q(r_{ij}), \\
q_1(r_{ij}) & = \frac{9}{5}q(r_{ij}), \\
q_2(r_{ij}) & = -\frac{2}{5}q(r_{ij}). \\
\end{aligned}
\end{equation}
In the case of a collinear state ($\textbf{s}_i \cdot \textbf{s}_j=1$), \textbf{Equation \ref{eq:Neel_energy}} is reduced to
\begin{equation}
\begin{aligned}
\mathcal{H}_{N\acute{e}el}^{\uparrow \uparrow}(\textbf{r},\textbf{s}) &= -\frac{1}{2}\sum_{i,j=1,i\neq j}^N \lbrace g(r_{ij}) + l(r_{ij})\left[ (\textbf{e}_{ij}\textbf{s}_i)^2-\frac{1}{3}\right]  \\ 
&+q(r_{ij})\left[(\textbf{e}_{ij}\textbf{s}_i)^4-\frac{6}{7}(\textbf{e}_{ij}\textbf{s}_i)^2+ \frac{3}{35}\right]\rbrace,
\end{aligned}
\label{eq:HeelFM}
\end{equation}
for FM ordered spins,\cite{PhysRevB.103.094437}  while assuming antiparallel pair spins ($\textbf{s}_i \cdot \textbf{s}_j=-1$) we obtain
\begin{equation}
\begin{aligned}
\mathcal{H}_{N\acute{e}el}^{\uparrow \downarrow}(\textbf{r},\textbf{s}) &= -\frac{1}{2}\sum_{i,j=1,i\neq j}^N \lbrace g(r_{ij}) - l(r_{ij})\left[ (\textbf{e}_{ij}\textbf{s}_i)^2-\frac{1}{3}\right]  \\ 
&+q(r_{ij})\left[\frac{13}{5}(\textbf{e}_{ij}\textbf{s}_i)^4+\frac{6}{7}(\textbf{e}_{ij}\textbf{s}_i)^2+ \frac{11}{35}\right]\rbrace,
\end{aligned}
\label{eq:HeelAFM}
\end{equation}
for AFM order.

 The dipole $l(r_{ij})$ and quadrupole $q(r_{ij})$ terms can describe the anisotropic effects induced by spin-orbit coupling like the anisotropic magnetostriction ($\lambda_{100}$ and $\lambda_{111}$) and magnetocrystalline anisotropy (MCA), respectively.\cite{PhysRevB.103.094437}  
 
 Since embedded atom method (EAM) potentials, commonly used in spin-lattice simulations, are either fitted to experimental or ab initio data, the influence of the 
 exchange interaction  is already silently incorporated in them. 
The term $g(r_{ij})$ can be used to shift ground state energy of the exchange interaction, for the sake of simplicity in the present model we do not include such offset energy, so that we set $g(r_{ij})=0$.

The spatial dependences of $J(r_{ij})$,  and   $l(r_{ij})$ and $q(r_{ij})$ are described using the Bethe-Slater curve $\Lambda(r_{ij})$, as implemented in the SPIN package of LAMMPS\cite{TRANCHIDA2018406}
\begin{equation}
\begin{aligned}
\Lambda(r_{ij}) & =  4\alpha_\Lambda \left(\frac{r_{ij}}{\delta_\Lambda}\right)^2 \left[1-\gamma_\Lambda\left(\frac{r_{ij}}{\delta_\Lambda}\right)^2\right] e^{-\left(\frac{r_{ij}}{\delta_\Lambda}\right)^2}\Theta(R_{c,\Lambda}-r_{ij}), 
\label{eq:BS_J_g_q}
\end{aligned}
\end{equation}
where $\Theta(R_{c,\Lambda}-r_{ij})$ is the Heaviside step function and the $R_{c,\Lambda}$ ($\Lambda=J,l,q$) are the cut-off radii. The main idea of the approach is to determine the parameters $\alpha_\Lambda$, $\gamma_\Lambda$, $\delta_\Lambda$ ($\Lambda=J,l,q$) in such a way that $J(r_{ij})$ reproduces the correct characteristic temperature  of transition from an ordered to a disordered magnetic state (Curie  or N\'{e}el  temperature) and the spontaneous volume magnetostriction $\omega_s$, $l(r_{ij})$ reproduces the anisotropic magnetostriction and $q(r_{ij})$ yields the MCA.

\subsection{The Bethe-Slater parameters of N\'{e}el  interaction for cubic antiferromagnets}
\label{section:Bethe-Slater}
\begin{figure}[h!]
\centering
\includegraphics[width=0.6\columnwidth ,angle=0]{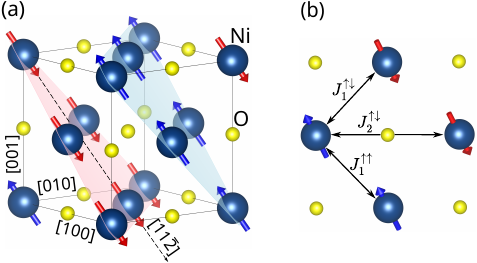}
\caption{(a) 
Crystal and magnetic structure of NiO. (b) The exchange interactions
between nearest neighbor and next-nearest-neighbor Ni sites.}
\label{fig:mag-order}
\end{figure}

Spin-lattice modeling of AFM is complicated. Unlike cubic FM, it requires modeling of various interactions between different neighbors. Part of them have co-aligned magnetic moments, and some have oppositely directed  ones, as it is apparent from the sketch of AFM on  \textbf{Figure \ref{fig:mag-order}}. Thus, in order to construct an accurate and general model, it is necessary to consider approaches for determining the parameters of the Bethe-Slater curves described above, i.e., $J(r_{ij})$, $l(r_{ij})$, and $q(r_{ij})$ for both AFM and FM oriented magnetic moments of the various neighboring Ni atoms.

The  generalized procedure to obtain the parameters of Bethe-Slater curves for cubic FM crystals\cite{PhysRevB.103.094437, book-Chikazumi} can be  expanded to the case of AFM cubic crystals, where it consists of the following  dependencies 
\begin{equation}
\begin{aligned}
\delta_\Lambda & = r_0, \\
\alpha_\Lambda &  = \frac{ e}{8}\left[2\Lambda(r_0) - r_0\frac{\partial \Lambda}{\partial r}\Big\vert_{r=r_0} \right], \\
\gamma_\Lambda &  = \frac{ r_0\frac{\partial \Lambda}{\partial r}\Big\vert_{r=r_0}}{r_0\frac{\partial \Lambda}{\partial r}\Big\vert_{r=r_0} - 2\Lambda(r_0)  },\\
\label{eq:BS_param}     
\end{aligned}
\end{equation}
where $e$ is Euler's number, and again $\Lambda=J,l,q$.  The cut-off radii $R_{c,\Lambda}$  should be {sufficiently large to include the nearest neighbors (NN) or next nearest neighbors (NNN) whose mutual interaction is being modeled, where $r_0$ is the distance to those neighbors. Parameters  $J(r_0)$, $l(r_0)$, $q(r_0)$ as well as derivatives $\frac{\partial J}{\partial r}\Big\vert_{r=r_0}$, $\frac{\partial l}{\partial r}\Big\vert_{r=r_0}$, $\frac{\partial q}{\partial r}\Big\vert_{r=r_0}$ are calculated with respect to $r_0$ and the relative magnetic moment orientations ($\uparrow \uparrow$ or $\uparrow \downarrow$). 

Due to the complexity of the NiO unit cell, it is convenient to choose carefully which magnetic pair interactions should be included in the spin-lattice model. Firstly, we consider magnetic interactions only between Ni atoms, which is reasonable since the magnetic moment of O is very small. Secondly, we notice that the Ni sublattice corresponds to a face-centered cubic (FCC), where 6  NN are parallel pair spins and the other 6 NN are antiparallel, see  Figure~
\ref{fig:mag-order}. This fact complicates  theoretical calculations of the Bethe-Slater parameters in \textbf{Equation \ref{eq:BS_param}}. On the other hand, the NNN of Ni correspond to the simple cubic (SC) structures where all pair spins are antiparallel, so that the derivation of the desired Bethe-Slater parameters in Equation \ref{eq:BS_param} is much easier. Hence, in the present model, we only include magnetic interactions (exchange and N\'{e}el terms) between Ni NNN. This choice can be also justified from a physical point of view for the exchange interactions, since the NNN exchange interaction ($J_2^{\uparrow \downarrow}$) is stronger than the NN ones ($J_1^{\uparrow \downarrow}$ and $J_1^{\uparrow \uparrow}$), see  \textbf{Table \ref{table:3}}.

The parameters of the Bethe-Slater curve for $J(r_{ij})$ are calculated to reproduce the desired N\'{e}el  temperature ($T_{N}$) and spontaneous volume magnetostriction ($\omega_s$). 
 From the analysis of the Mean Field Approximation (MFA) and N\'{e}el model\cite{book-Chikazumi} for an SC structure in an AFM state
it is found
\begin{equation}
\begin{aligned}
 J(r_0) = -\frac{k_BT_N}{2},\; r_0\frac{\partial J}{\partial r}\Big\vert_{r=r_0} = -\frac{\omega_s (C_{11}+2C_{12}) V_0}{ 3n},
\label{eq:J_dJafm}     
\end{aligned}
\end{equation}
where $k_B$ is the Boltzmann constant, $C_{11}$ and $C_{12}$ are the elastic constants, $r_0$ is the equilibrium distance to the NNN 
neighbors which is equal to the equilibrium lattice parameter $a_0$, and $n$ is the number of magnetic interacting atoms in the equilibrium volume $V_0$ that have been included in the model. For example, in the unit cell of NiO with volume $V_0=a_0^3$ we have $n=4$, see  Figure \ref{fig:mag-order} and \textbf{Figure \ref{fig:mag-subl}}. 

Similarly, the N\'{e}el dipole term $l(r_{ij})$ describes anisotropic magnetoelastic constants $b_1$ and $b_2$. Applying the N\'{e}el energy expression given by \textbf{Equation \ref{eq:HeelAFM}} for the SC case with AFM order we find
\begin{equation}
\begin{aligned}
 l(r_0) = \frac{V_0 b_2}{2n},\; r_0\frac{\partial l}{\partial r}\Big\vert_{r=r_0} = \frac{ V_0 b_1}{ n}.
\label{eq:l_dlafm}     
\end{aligned}
\end{equation}
Lastly, the N\'{e}el quadrupole term $q(r_{ij})$ simulates MCA in a cubic crystal. Using again Equation \ref{eq:HeelAFM} for the SC case with AFM order we obtain
\begin{equation}
\begin{aligned}
 q(r_0) =\frac{5 V_0 K_1}{26n},\; r_0\frac{\partial q}{\partial r}\Big\vert_{r=r_0} =\frac{15 V_0 K_1}{26n} \left[1- \frac{B}{K_1}\frac{\partial K_1}{\partial P}\right],
\label{eq:q_dqafm}     
\end{aligned}
\end{equation}
where $K_1$ is the first MCA constant, $B$ is the bulk modulus and $P$ is pressure. Note that these expressions are different to the SC case with FM order.\cite{PhysRevB.103.094437,book-Chikazumi}

\section{Spin-lattice model for $\textrm{NiO}$}
\subsection{Interatomic potential}
\label{section:int-potential}

Empirical interatomic potentials are designed to reproduce elastic properties in MD, the same as in SD-MD models. Thus choosing the correct potential is crucial.
However, such newly developed NiO potential has not been published yet, 
since both the elastic properties measured experimentally by various methods and those calculated from first principles can differ significantly (as will be shown later in Table \ref{table:3}). 
This can be partly explained by the fact that   
it is hard to separate the elastic contribution
from the influence of other effects in both experimental and theoretical results. 

In view of the above, in the present work, we propose examples of SD-MD model development based on two different potentials.  Each of them describes better a certain set of experimental and calculated elastic properties and thus is  more (or less) suitable for a description of particular effects/features. 

\subsubsection{ Born model  of ionic solids potential}
\label{section:int-potential-1}

As first potential for  molecular dynamics in NiO, we use the interionic potential model  proposed by Fisher and  Matsubara,\cite{Fisher-potential} based on the Born model of ionic solids. In their model, ions $i$ and $j$ interact with each other through long-range Coulombic interactions and short-range interactions that represent Pauli repulsions and van der Waal’s attractions. As a short-range term the Buckingham potential is used:
\begin{equation}
\phi (r_{ij})=A \exp{(-Br_{ij})}-\frac{C}{r_{ij}^6},
\end{equation}
where $A$, $B$ and $C$ are potential parameters particular to each ion–ion interaction (see \textbf{Table \ref{table:1}}).
\begin{table}[ht]
\caption{Buckingham potential parameters\cite{Fisher-potential}  used in the simulations}
\label{table:1}
\begin{tabular*}{0.95\columnwidth}{@{\extracolsep\fill}c|c|c|c}
\toprule
Interaction& $A$ (eV)&$B$ (\AA$^{-1}$)&$C$ (eV \AA$^{6}$)\\
\midrule
Ni$^{2+}-$ Ni$^{2+}$&0&1&0 \\
Ni$^{2+} -$ O$^{2-}$&754.92&3.05157&0 \\  
O$^{2-} -$ O$^{2-}$&22764.3&6.71141& 27.89\\
\midrule
\end{tabular*}
\end{table}

At the initial stage, it is convenient to first consider elastic  properties obtained by using only the interatomic potential and find the equilibrium volume, bulk modulus and elastic constants.
Thus, computing the energy of the fcc unit cell of NiO for different volumes using the open-source code LAMMPS\cite{thompson2022lammps} and then fitting the resulting data to the Murnaghan equation of state (EOS)\cite{PhysRevB.28.5480} gives the equilibrium state with a cell volume $V_0=  73.367$ \r{A}$^3$ (lattice constant $a_0=a_0^{NM}=4.18633$ \r{A}) and a bulk modulus $B=209.83$ GPa. We have verified that for the obtained equilibrium value the pressure does not exceed $2.5\times 10^{-3}$ GPa as can be seen from  \textbf{Figure \ref{fig:En-vs-V}}.
The elastic constants obtained from the potential are $C_{11}=287$, $C_{12}=171$, $C_{44}=171$ GPa, relatively close to those calculated by DFT\cite{MaterialsProject} and experimentally obtained\cite{10.1121/1.1913005} (see Table~\ref{table:3}). 
Although the obtained $C_{ij}$ do not completely coincide with the experimental ones, it is quite  possible that a better agreement with the experimental data might naturally appear later in SD-MD simulations when the magnetic properties of the crystal and the resulting small lattice distortions are taken into account.





Hereafter, we will denote the SD-MD model that uses this potential as SD-MD 1.

\begin{figure}[h!]
\centering
\includegraphics[width=0.6\columnwidth ,angle=0]{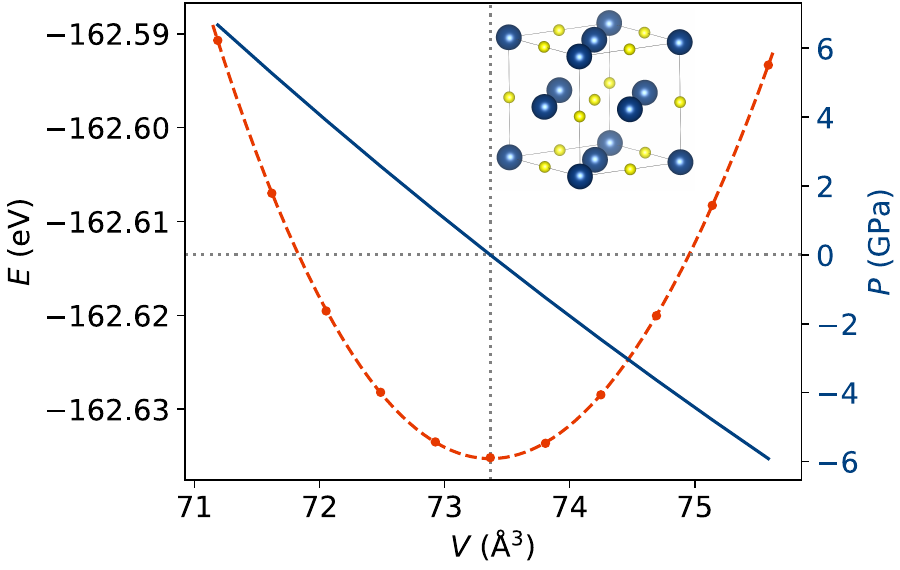}
\caption{Total energy and pressure as a function of the cell volume changes.
The red dots depict the results obtained from the MD simulation in LAMMPS by just including the interionic potential modeled by Fisher and Matsubara\cite{Fisher-potential}  and the dashed line corresponds to the EOS fit. The blue 
line shows  the pressure.  }
\label{fig:En-vs-V}
\end{figure}

\subsubsection{
Reference-free modified embedded atom method interatomic potential \label{computational_details} }

As an alternative to the previous interatomic potential, we construct an ab initio data-based  one within the reference-free modified embedded atom method (RF-MEAM)\cite{r15_Duff_CompPhysComm,r22_Slooter_JPCM}. 
The potential is fitted by means of the MEAMfit2 code\cite{r15_Duff_CompPhysComm} to a dataset of DFT results related to 45 distorted NiO crystal structures. They are  calculated via the plane-wave based Vienna Ab initio Simulation Package (VASP)\cite{vasp_r96_prb,vasp_r99_prb}, considering the projector-augmented-wave method\cite{PhysRevB.50.17953} pseudo-potentials. Non-collinear magnetic calculations, including spin-orbit coupling, were performed in the generalized gradient approximation of Perdew-Burke-Ernzerhof\cite{pbe1} including 16 Ni valence electrons 3p$^{6}$, 4d$^{9}$, 4s$^{1}$ and 6 O valence electrons 2s$^{2}$, 2p$^{4}$, with an energy cut-off  for the plane waves of 520 eV and an automatically generated $k$-mesh scheme with $R_{k}$~=~40 (5x5x5 k-mesh), where the distorted structures were generated via the AELAS package\cite{AELAS_r2017}. To fit the  experimental behavior, the Hubbard  correction $U=5$~eV for Ni d-states in the {Dudarev} approach\cite{Dudarev_r98_prb57} is applied.

The advantage of such type of interatomic potentials is  the possibility to develop custom potentials with comparatively good consistency between DFT and simulated results for the studied systems.
Similar calculations to those in Section \ref{section:int-potential-1}, including only the present interatomic potential, give by EOS fitting  the lattice constant $a_0^{NM}=4.24038$~\r{A}, where the  pressure at the obtained equilibrium does not exceed $2.6\times 10^{-3}$ GPa (see  Figure \ref{fig:En-vs-V2}). The derived elastic constants  $C_{11}=320$, $C_{12}=114$, 
$C_{44}=82$ GPa and a bulk modulus $B=182.7$ GPa are close to those obtained by DFT (Table \ref{table:3}). 
 
\begin{figure}[h!]
\centering
\includegraphics[width=0.6\columnwidth ,angle=0]{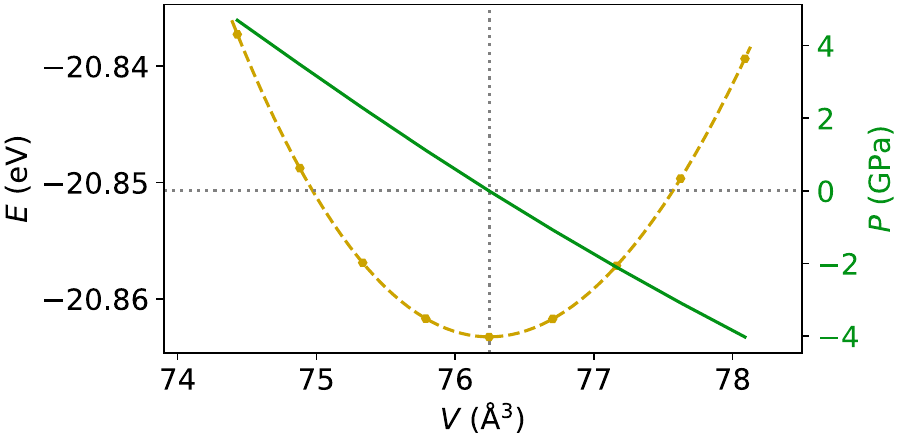}
\caption{Total energy as a function of the cell volume changes fitted by EOS for the case of  NiO MD simulations performed with the RF-MEAM potential.
The green 
line shows the pressure.  }
\label{fig:En-vs-V2}
\end{figure}
We will denote the SD-MD model that uses this RF-MEAM potential as SD-MD 2.

\subsection{Magnetic ordering}
NiO has a simple rock salt structure (space group Fm$\bar3$m) above the N\'eel temperature $T_N = 523$ K. Below $T_N$, the spins of the Ni$^{2+}$ ions are ordered ferromagnetically in \{111\} planes where they lie along $\langle11\bar2\rangle$ axes. In adjacent \{111\} planes the sign of the ferromagnetic order is opposite resulting in a type-II fcc AFM compound.\cite{PhysRevLett.105.077402}
As it can be seen from  \textbf{Figure \ref{fig:mag-subl}}, if only Ni ions are taken into account then NiO has an fcc structure with a bulk unit cell parameter of $a_0$.
However, strong superexchange and resulting AFM magnetic ordering make it more complex for modeling and analytical description. 
Thus, to reveal the magnetic order, a 2x2x2 supercell consisting of 64 atoms is required. In such magnetic cell with lattice parameter $a=2a_0$, 32  Ni$^{2+}$ atomic moments ($\mu=1.9$ $\mu_B$) are distributed among 8 magnetic sublattices, being paired to generate 4 antiferromagnetic submotifs. The magnetic sublattices are shown on  Figure \ref{fig:mag-subl}.


\begin{figure}[h!]
\centering
\includegraphics[width=0.4\columnwidth ,angle=0]{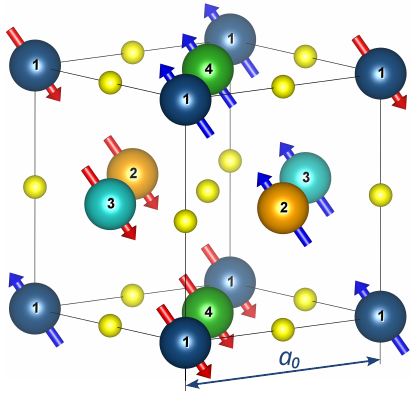}
\caption{Definition of the SC Ni$^{2+}$ sublattices in NiO. Each of the 4 SC 
sublattices, in turn, consists of two sublattices with spin up and spin down.}
\label{fig:mag-subl}
\end{figure}

\subsection{Exchange interaction}
As mentioned earlier,  in the exchange interaction, and due to strong superexchange in NiO, the contribution from the AFM coupled second nearest neighbors prevails  over the others (i.e., $J_2^{\uparrow \downarrow} \gg J_1^{\uparrow \uparrow}, J_1^{\uparrow \downarrow}$), 
and the description can be simplified by restricting the treatment only to the NNNs. Thus, in our model, only  the exchange interactions $J_2^{\uparrow \downarrow}$, between Ni atoms inside the 4 simple cubic (SC) sublattices (Figure \ref{fig:mag-subl}), are considered, whereas  the inter-sublattice interactions   $J_1^{\uparrow \uparrow}$ and $J_1^{\uparrow \downarrow}$ are neglected.

To describe $J(r)$, we need to specify the input parameters $T_N$, $C_{11}$, $C_{12}$, $C_{44}$ and $\omega_s$ see \textbf{Equation \ref{eq:J_dJafm}}.
They can be taken either from experiment  or DFT calculations. 
Therefore, we use the known experimental value of the N\'eel temperature, namely, 523 K\cite{PhysRevB.6.3447} and the elastic constants as determined by the interionic potential (Section \ref{section:int-potential}, as shown in Table \ref{table:3}). 
The volume magnetostriction at zero temperature can be estimated as $\omega_s=(a_{AFM}^3 - a_{PM}^3)/a_{PM}^3=-0.00143$, based on the DFT calculation of the lattice constants for the AFM ($a_{AFM}$) and paramagnetic ($a_{PM}$) states given by Plummer \textit{et al.}\cite{10.1063/5.0246100}


As other required parameters for the Bethe-Slater curve $J(r_{ij})$ parametrization according to Equation 
\ref{eq:BS_param} and \ref{eq:J_dJafm}, we take the distance between second nearest AFM oriented neighbors  $r_0=a_0^{NM}$, obtained in Section \ref{section:int-potential} from the potential for the non-magnetic (NM) case, i.e., without taking into account spins, and the total number of atoms in the equilibrium volume $V_0=r_0^3$, which is $n=4$ for an fcc unit cell.
The resulting parameters $R_{c,J}$, $\alpha_J$ and $\delta_J$ are given in \textbf{Table \ref{table:2}}.

\begin{table}[ht]
\caption{Parameters of the Bethe-Slater curves using the SD-MD model for NiO}
\label{table:2}
\begin{tabular*}{0.9\columnwidth}{@{\extracolsep\fill}c|c c}
\toprule
SD-MD parameters & SD-MD 1 & SD-MD 2\\
\midrule
$R_{c,J}$ (\r{A})& 4.5 & 4.5\\
$\alpha_J$ (meV/atom)&  -26.98139& -25.87886 \\  
$\gamma_J$ &  0.4324364 & 0.4082562\\
$\delta_J$ (\r{A})& 4.18633&4.24038\\
$R_{c,l}$ (\r{A}) & 4.5  & 4.5 \\
$\alpha_l$ ($\mu$eV/atom)&  29.08110& 30.22210\\  
$\gamma_l$ &  -1.502436 & -1.502436\\
$\delta_l$ (\r{A})& 4.18633 & 4.24038\\
$R_{c,q}$ (\r{A}) & 4.5 & 4.5 \\
$\alpha_q$ ($\mu$eV/atom)& 3.804386 & 3.342015\\  
$\gamma_q$ &  0.5461027 & 0.4630328\\
$\delta_q$ (\r{A})& 4.18633 & 4.24038\\
\midrule
\end{tabular*}
\end{table}

\subsection{Ab initio  calculations of input parameters for N\'{e}el energy parametrization}
\label{section:ab-initio}

Firstly, we have computed the $C_{ij}$ by means of an ab initio  approach using the AELAS code\cite{AELAS_r2017} interfaced with VASP\cite{vasp1} for a lattice parameter $a_{lat}=4.22$ \r{A}, leading to $C_{11}=340$ GPa, $C_{12}=116$ GPa, $C_{44}=85$ GPa and bulk modulus $B=190$ GPa, using the same parameters as above and 13 distortions with maximal relative size of $\pm$0.018. 
This result is relatively close to that experimentally found by M. Grimsditch \textit{et al}.\cite{grimsditch1994brillouin}

To calculate the Bethe-Slater parameters for the dipole term we need to know the values of the anisotropic magnetoelastic constants $b_1$ and $b_2$. To determine them from ab initio calculations  we use the MAELAS code with mode 2 based on the strain-energy method\cite{NIEVES2022108197} in combination with VASP for electronic structure calculations including spin-orbit coupling (SOC).\cite{vasp1}
The idea of the method is to subtract the total energy from two different magnetization directions for a deformed unit cell in such a way that we can get the $i$-th anisotropic magnetoelastic constant $b_i$ from a linear fitting of the energy versus strain data.\cite{ NIEVES2022108197} Thus, for a cubic crystal, such linear dependences have the form
\begin{equation}
\begin{aligned}
\frac{1}{V_0}(E_{[100]}(\varepsilon_{xx})-E_{[110]}(\varepsilon_{xx}))&=\frac{1}{2}b_1\varepsilon_{xx}-\frac{1}{4}K_1,\\
\frac{1}{V_0}(E_{[110]}(\varepsilon_{xy})-E_{[\bar{1}10]}(\varepsilon_{xy}))&=2b_2\varepsilon_{xy},
\label{eq:b1-b2}   
\end{aligned}
\end{equation}
being $V_0$ the equilibrium volume of the magnetic supercell, where AFM order is used. 
In order to do so, we have increased the number of k-points to 216 in the half Brillouin zone to accurately capture the total energy vs. strain for both spin directions, while other computational details remain the same as in Section \ref{computational_details}.

From the linear fitting of the energy versus strain data 
we obtain $b_1$ = 1.12  MPa and $b_2$= 1.87 MPa. Then we used them to calculate the magnetostrictive constants as $\lambda_{100}=-2b_1/3(C_{11}-C_{12})=-3.343\times10^{-6}$ and $\lambda_{111}=-b_2/3C_{44}=-7.337\times10^{-6}$. 
These results give the same sign, but smaller magnitude, as those from Phillips \textit{et al.}\cite{PhysRev.153.616} ($\lambda_{100}=-1.45\times10^{-4}$,  $\lambda_{111}=-0.79\times10^{-4}$), derived from measurements of the crystal-field tensor of NiO and the elastic constants of MgO.

Next, we use the same VASP settings as for $b_1$ and $b_2$ to find the pressure dependence of the magnetocrystalline anisotropy constant $K_1(P)$. 
To do so, we evaluate the total energy $E$ with AFM order along crystallographic directions [100] and [110], at different volumes $V$ of the $2\times2\times2$ supercell, and compute $K_1$ using 
\begin{equation}
    K_1(V)=\frac{4[E_{[110]}(V)-E_{[100]}(V)]}{V},
    \label{eq:K1}
\end{equation}
where for each volume we also compute the corresponding pressure $P$.
The pressure dependence of $K_1$ in the low pressure regime, where $ \zeta P \ll 1 $ (here for simplicity we use the notation $\frac{1}{K_1}\frac{\partial K_1}{\partial P}=\zeta$)  follows approximately a linear law\cite{SAWAOKA1975267, PhysRevB.103.094437} as follows
\begin{equation}
\frac{K_1(P)}{K_1(0)}\approx1+ \zeta P +O(P^2).
\label{eq:K1-P}     
\end{equation}
Thus, fitting the data to the above \textbf{Equation \ref{eq:K1-P}} allows us to obtain the necessary input parameters $K_1(0)$ and $\frac{1}{K_1}\frac{\partial K_1}{\partial P}$ for the SD-MD. 
This procedure gives the values $K_1=115.42$ kJ m$^{-3}$, which is in excellent agreement with the value given 
by Schr\"on \textit{et al.}\cite{PhysRevB.86.115134},
and $\frac{1}{K_1}\frac{\partial K_1}{\partial P}=0.00862$ GPa$^{-1}$, for which there is no available data in the literature. 


\subsection{N\'{e}el energy}
Similar to what was done previously,  only the interactions between second nearest, AFM ordered, neighbors are taken into account to model the Néel energy. All other pair interactions are neglected. 

In order to parameterize  $l(r)$ and $q(r)$ terms  in Equation \ref{eq:HeelAFM} with \textbf{Equation \ref{eq:l_dlafm}} and \textbf{\ref{eq:q_dqafm}}, respectively, we use the values of $b_1$, $b_2$, $K_1$ and $\frac{1}{K_1}\frac{\partial K_1}{\partial P}$ obtained in Section \ref{section:ab-initio} and analogously to the parameterization of the $J(r)$ interaction at the equilibrium volume $V_0=r_0^3$, we set the distance between second nearest neighbors to $r_0=a_0^{NM}$ and the number of atoms in the equilibrium volume to $n=4$. 
Although, in general, the inclusion of the N\'{e}el energy affects $r_0$, this effect is relatively small, which allows us to use the same value as in the parameterization of the exchange interaction. We also used the bulk modulus $B$ found by means of the interatomic potentials presented in Section \ref{section:int-potential}. The obtained Bethe-Slater parameters for $l(r)$ and $q(r)$ are shown in Table \ref{table:2}.

\section{Results}
\subsection{Tests of spin-lattice model}
\subsubsection{Volume magnetostriction}
The exchange-induced volume magnetostriction can be calculated from the SD-MD model as
$\omega_s=\frac{V_0^{AFM} - V_0^{PM}}{V_0^{PM}}$, where $V_0^{AFM}$ and $V_0^{PM}$ are the equilibrium volumes of the antiferromagnetic and paramagnetic cells, respectively.
In the simulations, 
we use a supercell with 85184 atoms and set the paramagnetic state by using a random orientation of the spins. To find the equilibrium volumes, we used an energy versus volume curve fitting by means of a Murnaghan EOS\cite{PhysRevB.28.5480}.

On  \textbf{Figure \ref{fig:ws}} we show the result obtained from SD-MD simulations which give a volume magnetostriction value of $-0.00149$ for SD-MD 1 and $-0.00136$ for SD-MD 2 models, respectively, thus demonstrating a good fit of the models to the magnetostriction value $\omega_s=-0.00143$ embedded in them.  

In the graphs we show the volume $V=a_0^3$ which is convenient for comparison with the NM state, although we point out that the magnetic cell used in our simulations is 8 times bigger. 
It is also expected that increasing the size of the PM supercell will give an $\omega_s$ value closer to that used as an input parameter in the model.

\begin{figure}[h!]
\centering
\includegraphics[width=0.6\columnwidth ,angle=0]{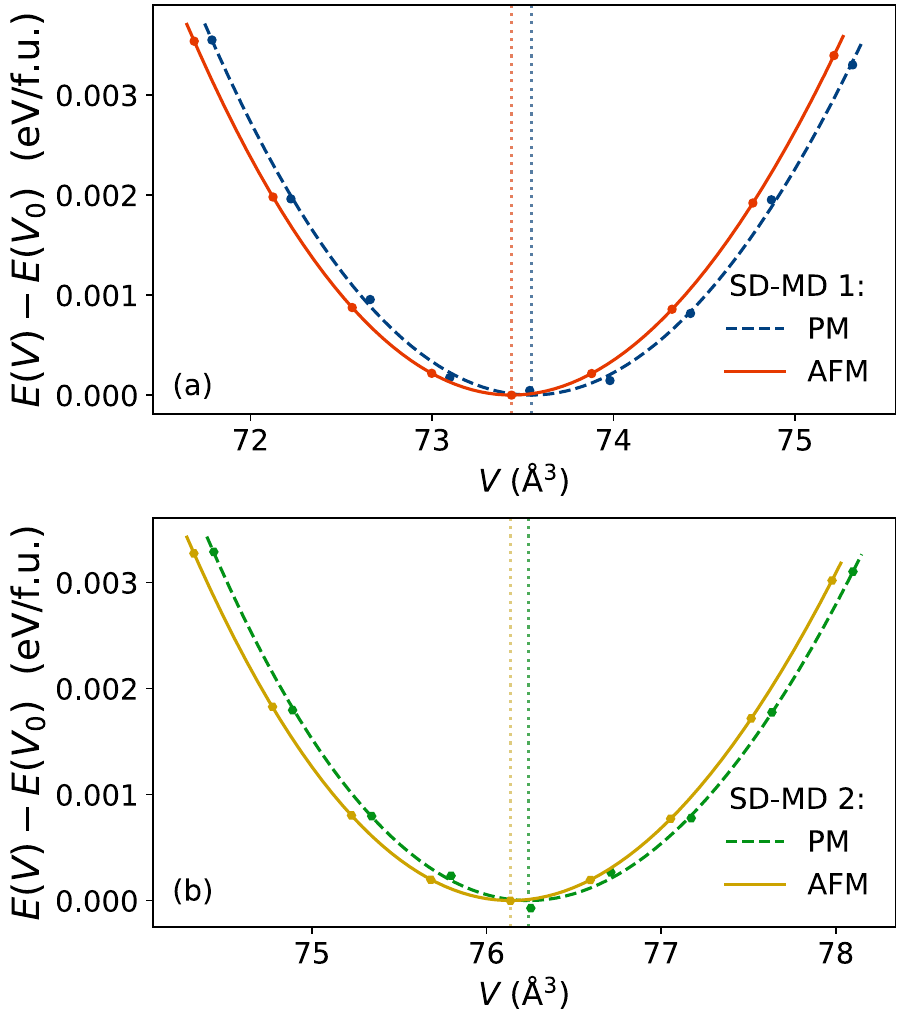}
\caption{Volume magnetostriction in SD-MD models. Equilibrium volumes for AFM and PM states were derived from fitting simulation data by EOS giving the volume magnetostriction constant $\omega_s=-0.00149$ in (a) SD-MD 1 and $\omega_s=-0.00136$ in (b) SD-MD 2. The dotted lines show the equilibrium volumes of the corresponding AFM and PM states.
The energy difference between the PM and AFM ordered equilibrium states is $E(V_0^{PM})-E(V_0^{AFM})=0.03378$ (eV/f.u.) in both SD-MD models.}
\label{fig:ws}
\end{figure}

As discussed in Section \ref{section:Hamiltonian}, since the exchange energy offset procedure was not applied, a small pressure may remain in the equilibrium state in this model. Thus, in the case of SD-MD 1, for the AFM state, the equilibrium lattice parameter found from the EOS fit $a_0^{AFM}=4.18769$ \r{A} and the pressure does not exceed 0.3 GPa. For the PM state these values are $a_0^{PM}=4.18978$ \r{A} and $1\times 10^{-3}$ GPa.
In the case of SD-MD 2 these values are  $a_0^{AFM}=4.23837$ \r{A}, where the pressure does not exceed  $2.72\times10^{-3}$ GPa and $a_0^{PM}=4.24029$ \r{A} with the pressure less than $2\times 10^{-4}$ GPa.

\begin{table*}[ht]
\caption{Parameters of NiO with SD-MD and parameters measured experimentally or calculated in other sources.
}
\label{table:3}
\begin{tabular*}{\textwidth}{@{\extracolsep\fill}c|c c c c c c}
\toprule
Parameters & SD-MD 1 & SD-MD 2  &Calc. (present work) & Calc. &Expt. &Expt.\\
\midrule
$a_0^{NM}$ (\r{A})& 4.18633& 4.24038& & & & \\
$a_0^{AFM}$ (\r{A})& 4.18769 & 4.23837 & 4.22 & 4.190 \cite{10.1063/5.0246100}& 4. 1705 \cite{PhysRevB.6.3447}&\\
$a_0^{PM}$ (\r{A})&  4.18978 & 4.24029 && 4.192 \cite{10.1063/5.0246100} & &\\
$J_2^{\uparrow \downarrow}$ (meV)& -22.5 & -22.5 & & -14.0 \cite{PhysRevB.88.134427} & -19.01 \cite{PhysRevB.6.3447}&\\
$J_1^{\uparrow \uparrow}$, $J_1^{\uparrow \downarrow}$ (meV)& 0 & 0 &  & 1.2 \cite{PhysRevB.88.134427}& 1.37 \cite{PhysRevB.6.3447}&\\
$\omega_s$& -0.00149 & -0.00136 && -0.00143 \cite{10.1063/5.0246100}& &\\
$C_{11}$ (GPa)& 287& 320 & 340 & 274 \cite{MaterialsProject}& 270 \cite{10.1121/1.1913005} & 358 \cite{grimsditch1994brillouin}\\
$C_{12}$ (GPa)& 171& 114 & 116& 170 \cite{MaterialsProject}& 125 \cite{10.1121/1.1913005} &134 \cite{grimsditch1994brillouin}\\
$C_{44}$ (GPa)& 171& 82 & 85 & 83 \cite{MaterialsProject}& 105 \cite{10.1121/1.1913005} &93 \cite{grimsditch1994brillouin}\\
$B$ (GPa)& 209.8& 182.7 & 190 & 194 \cite{10.1063/5.0246100} & 173.8 \cite{10.1121/1.1913005} & $180-220$ \cite{PhysRevB.84.115114}\\
$b_{1}$ (MPa)& 1.38 & 1.36 & 1.12 && &\\
$b_{2}$ (MPa)& 1.72 & 1.75 & 1.87 && &\\
$\lambda_{100}$ ($\times 10^{-4}$)& -0.079 & -0.036 &  -0.033 &-1.45 \cite{PhysRev.153.616}& &\\
$\lambda_{111}$ ($\times 10^{-4}$)& -0.033 & -0.076 & -0.073 &-0.79 \cite{PhysRev.153.616}& &\\
$K_1$ (kJ m$^{-3}$)& 115.42& 115.42 &115.42 &113.5 \cite{PhysRevB.86.115134}& &\\
$\frac{1}{K_1}\frac{\partial K_1}{\partial P}$ (GPa$^{-1}$)& 0.00826 & 0.00816 &0.00862 && &\\
\midrule
\end{tabular*}
\end{table*}

\subsubsection{Magnetocrystaline anisotropy}
\label{subsection:MCA}
The correctness of the MCA in the model is checked by specifying the directions of the magnetic moments (leaving unchanged their relative antiferromagnetic order with respect to their neighbors) in the equilibrium volume as parallel to [110] and [100] directions, followed by comparing the energies of these states as $K_1=4(E_{[110]}-E_{[100]})/V_0$.
 The obtained value of the anisotropy constant $K_1=115.42$ kJ m$^{-3}$ for both SD-MD 1 and SD-MD 2 is in exact agreement with that calculated in Section \ref{section:ab-initio}.
\begin{figure}[h!]
\centering
\includegraphics[width=0.6\columnwidth,angle=0]{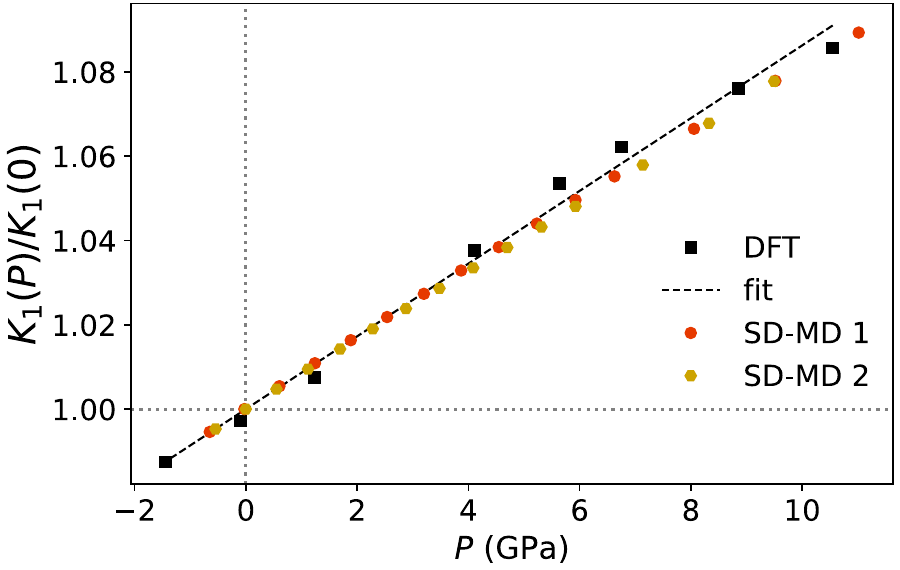}
\caption{Calculation of the hydrostatic pressure ($P$) effect on MCA ($K_1$) for NiO. The black squares represent the result of the DFT calculations, the dashed line shows their fit to Equation \ref{eq:K1-P}, the red circles and yellow hexagons correspond to the results from the SD-MD simulations.}
\label{fig:K1-P}
\end{figure}

As it was shown on the basis of cubic FM by Nieves \textit{et al.}\cite{PhysRevB.103.094437} the effect of the hydrostatic pressure on MCA in the SD-MD model might be verified by analyzing the behavior of $K_1(P)$. 
The compliance or non-compliance with the linear law given by \textbf{Equation \ref{eq:K1-P}} can be easily detected in the data simulated with the SD-MD model.
In \textbf{Figure \ref{fig:K1-P}}  a comparison of the results obtained from DFT and SD-MD simulations is shown. Fitting the SD-MD simulations data with Equation \ref{eq:K1-P}  gives a values of $\frac{1}{K_1}\frac{\partial K_1}{\partial P}$  
equal to 0.00826 GPa$^{-1}$ for SD-MD 1 and 0.00816 GPa$^{-1}$ for SD-MD 2, which agrees well with the value of 0.00862 GPa$^{-1}$ for the DFT data fit.

\subsubsection{Anisotropic magnetostriction}

To further verify the magnetostrictive behavior in the proposed model, we use the interface between the program MAELAS and LAMMPS\cite{github}, slightly modified for the NiO case, which allows us to calculate the anisotropic magnetostrictive constants from the used SD-MD models.
Thus, following the general method described in Section \ref{section:ab-initio},  
SD-MD simulations are used to obtain the energy versus strain data and to find the magnetoelastic constants $b_1$ and $b_2$ from them.
This procedure allows  to check how accurately the SD-MD model reproduces the values of $b_1$ and $b_2$ that were obtained from ab initio calculations in Section \ref{section:ab-initio} and used in the model as input parameters.

\begin{figure}[h!]
\centering
\includegraphics[width=0.6\columnwidth ,angle=0]{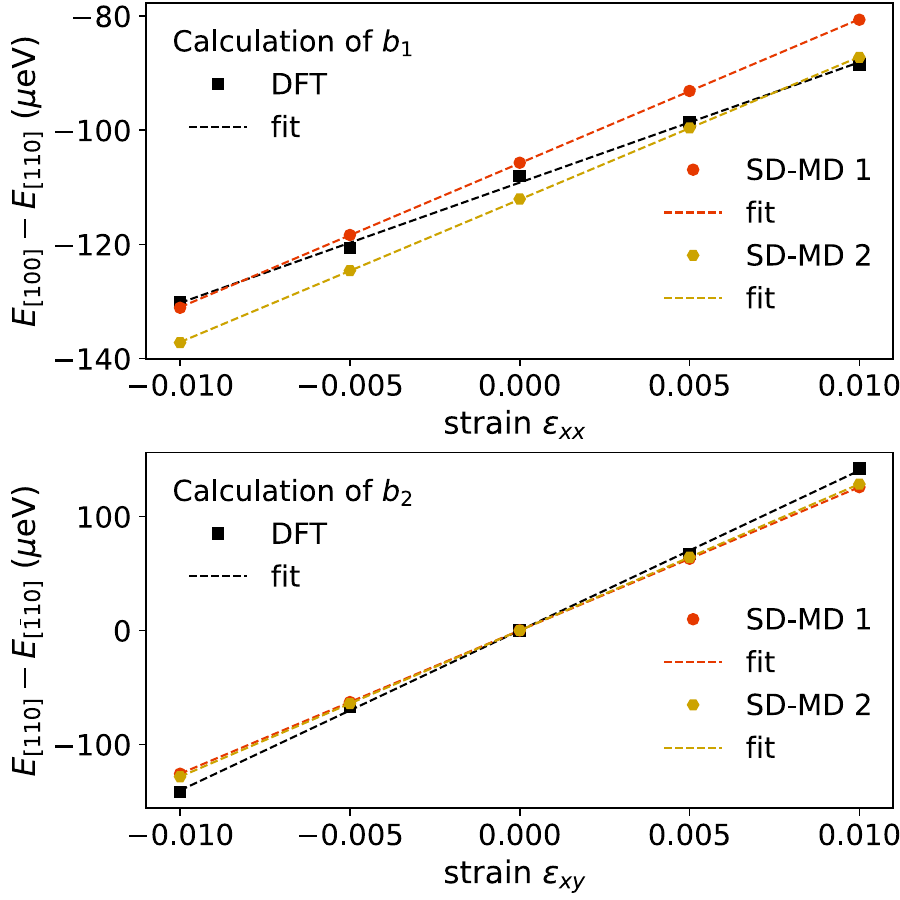}
\caption{Calculation of magnetoelastic constants $b_1$ and $b_2$ through a linear fitting of the energy versus strain data. Comparison of the results obtained with DFT and SD-MD.}
\label{fig:b1-b2}
\end{figure}

 \textbf{Figure \ref{fig:b1-b2}} shows the energy versus strain dependencies  obtained using  MAELAS for both DFT and SD-MD models.
The figure shows a good agreement between both models, as the observed small deviations in values lie well within the range of accuracy of the methods.



\subsection{Discussion}

As can be seen in the previous sections, the proposed methodology for creating spin-lattice models provides the correct magnetoelastic behavior for the both given examples of cubic AFM NiO. The results obtained from DFT calculations here are generally in good agreement with experimental data and calculations done by other scientists, except for magnetoelastic  constants $b_1$ and $b_2$, which give  $\lambda_{100}$ and $\lambda_{111}$  noticeably smaller than those calculated by Phillips \textit{et al.},\cite{PhysRev.153.616} where those constants for NiO were estimated using  the elastic constants of MgO. This situation does not change the effectiveness of the proposed spin-lattice model, but raises the question of additional research using modern methods to clarify the exact values of magnetostrictive constants in this material.

To increase the accuracy of the model, a recursive approach can be used, where the Bethe-Slater curve parameters are first obtained approximately from the non-magnetic model, and then the values of $a_0$, $C_{11}$, $C_{12}$, $C_{44}$ and $B$, obtained from simulations with included magnetic properties, are used as input data to find more accurate parameters $R_{c,k}$, $\alpha_k$, $\gamma_k$, $\delta_k$ ($k=J,l,q$). 

Another advantage of the proposed model is that it benefits from a natural integration of long-range dipolar interactions that are included in the SPIN package\cite{TRANCHIDA2018406} and may be required in the case of studying domain formation and shape effects in AFM.


It should also be noted that one of the challenges in modeling antiferromagnetic systems is related to the fact that to perform spin-lattice simulations, reliable interatomic potentials are required to ensure realistic behavior. The number of publicly available potentials is limited and most of them have been designed so far for single elements or binary systems. Therefore, it is desirable to develop custom potentials for the systems of interest. As a result, the present work provides a beautiful example of such creation and use of a custom interatomic RF-MEAM potential based on fitting our in-house DFT calculations and on a comparison of the results obtained with those found from using those available for the NiO potential based on the Born model of ionic solids. The results from such a comparison are shown in Table \ref{table:3}, where it can be seen that the new developed RF-MEAM potential avoids the general disadvantage of the Born model potentials associated with the quantity of $C_{44}$\cite{Lewis_1985} and thus leads to a better agreement in magnetostrictive coefficients $\lambda_{100}$ and $\lambda_{111}$.


\section{Conclusions}


This work represents a first attempt at developing interatomic potentials that include magnetic properties in cubic symmetry antiferromagnets. As a proof of concept, the methodology was successfully used for the case of NiO at zero temperature, allowing the study of magnetic effects in large-scale metal oxides through molecular dynamics simulations, which is particularly relevant due to the great technological interest of this type of materials. The results shown should be interpreted with caution in the absence of more detailed studies at finite temperature, but they are very promising given the large number of possibilities they offer. For example, in the particular case of NiO, it could help to better understand some of its properties, such as its easy axis and the influence of magnetoelastic effects on it. Furthermore, among the possible future applications of these models, we can highlight studies on magnetoelastic and magnetoelectric coupling phenomena, effects associated with magnetic fields, shape effects in magnetic nanostructures, and magnon-phonon dynamics.

\section*{Acknowledgement}
This work was supported by the Ministry of Education, Youth and Sports of the Czech Republic through the e-INFRA CZ (ID:90254) and by project QM4ST (CZ.02.01.01/00/22\_008/0004572). P. N. acknowledges support by grant MU-23-BG22/00168 funded by The Ministry of Universities of Spain. J. S. acknowledges GAČR project No. 24-11388I, whereas DL and IK project No. 25-14529L of the Grant Agency of Czech Republic. R.I. acknowledges funding from the project BETMASFUS, grant PID2023-149089OB-I00, Ministerio de Ciencia, Innovación y Universidades (Spain), the EIT-RM/EU project ExpSkills-REM, grant number UE-22-EXPSKILLS-21104, the Spanish AEI Project NEXPECH-2, grant number MCINN-24-PCI2024-153437 and the Agencia de Ciencia, Competitividad Empresarial e Innovación Asturiana (Sekuens) project MAGNES, SEK-25-GRU-GIC-24-113. Support provided by the Eu-MACE and EuMINe COST Actions CA22123 and CA22143, respectively, is gratefully acknowledged as well.

\bibliography{magnelast.bib}


\begin{figure}
	\textbf{Table of Contents}\\
	\medskip
	\includegraphics{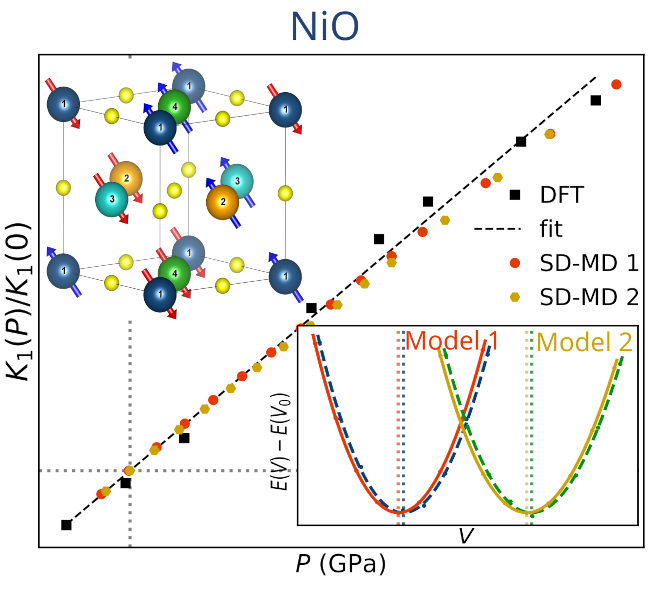}
	\medskip
	\caption*{Interatomic potentials are key to performing large-scale molecular dynamics simulations devoted to the study of materials. For magnetic materials, potentials must include magnetic features to achieve a more complete description of these materials. This work presents a methodology for incorporating magnetic properties into interatomic potentials for antiferromagnetic materials with cubic symmetry, and demonstrates an example of its application for NiO.}
\end{figure}

\end{document}